# One-dimensional phosphorus chain and two-dimensional blue phosphorene grown on Au(111) by molecular-beam epitaxy

Jin-Peng Xu[1], Jun-Qiu Zhang [1], Hao Tian[1,2], Hu Xu[2], Wingkin Ho[1] and Maohai Xie[1*]

[1]*Department of Physics, Hong Kong University, Hong Kong, China*
[2]*Department of Physics, Southern University of Science and Technology, Shenzhen, 518055, China*

**Abstract**

Single layer (SL) phosphorus (phosphorene) has drawn considerable research attention recently as a two-dimensional (2D) material for application promises. It is a semiconductor showing superior transport and optical properties. Few-layer or SL black phosphorus has been successfully isolated by exfoliation from bulk crystals and extensively studied thereof for its electronic and optical properties. Blue phosphorus (blueP), an allotrope of black phosphorus where atoms are arranged in a more flat atomic configuration, has been recently suggested by theory to exist in the SL form on some substrates. In this work, we report the formation of a blueP-like epilayer on Au(111) by molecular-beam epitaxy. In particular, we uncover by scanning tunneling microscopy (STM) one-dimensional (1D) atomic chains at low coverage, which develop into more compact islands or patches of $(\sqrt{3} \times \sqrt{3})R30°$ structure with increasing coverage before blueP-like islands nucleate and grow. We also note an interesting growth characteristic where the $(\sqrt{3} \times \sqrt{3})R30°$ surface at intermediate coverage tends to phase-separate into locally low-coverage 1D chain and high-coverage blueP-like structures, respectively. This experiment thus not only lends a support of the recently proposed half-layer by half-layer (HLBHL) growth mechanism but also reveals the kinetic details of blueP growth processes.

*Author of correspondence, Email: mhxie@hku.hk

After the success of graphene, extensive research efforts have recently been devoted to other two-dimensional (2D) systems. Examples include silicene [1,2], stanene [3,4], borophene [5,6], group-III and transition metal dichalcogenides [7-12]. Owing to the superior transport and optical properties, few-layer and single-layer (SL) black phosphorus (P) have come under the spotlight lately [13-16]. It represents one of the latest members in the expanding 2D family. Blue phosphorus (blueP), an allotrope of black P but arranged in a more flat atomic configuration and having similar appealing properties, has recently been suggested to exist in the SL form on some substrates [17-20]. Blue phosphorene, as is called, is a semiconductor with the fundamental gap near 2 eV [17]. While black phosphorene has been experimentally investigated for the exfoliated samples, studies of epitaxial black and blue phosphorene remain scarce.

The typical methods of obtaining few-layer or SL materials include exfoliation, chemical vapor deposition and molecular-beam epitaxy (MBE). The latter has been shown advantageous in producing many designer 2D materials such as silicene [2], stanene [4], and lately tellurene [21,22]. It however remains challenging to use MBE to grow phosphorene. A recent theoretical study has suggested that SL blueP can be stabilized on GaN(0001) via the unusual half-layer by half-layer (HLBHL) growth mechanism [19]. There is thus a high hope that epitaxial phosphorene can be obtained by MBE and if so it would accelerate both fundamental and practical researches of phosphorene films.

In this Letter, we report the growth of phosphorus by MBE. In particular, we reveal the HLBHL growth characteristics of P on Au(111), conforming to the recent theoretical prediction [19]. By using scanning tunneling microscopy (STM), we further uncover a new one-dimensional (1D) chain structure, which evolves into the $(\sqrt{3} \times \sqrt{3})R30°$ compact islands or patches with increasing coverage before blueP-like islands nucleate and grow. The latter manifest by the $(5 \times 5)$ reconstruction, similar to that reported earlier [18,20]. More interestingly, we discover that the $(\sqrt{3} \times \sqrt{3})R30°$ phase at intermediate coverage gives way to the "phase-separated" surfaces consisting of locally low-coverage 1D chains and high-coverage blueP-like islands coexisted side-by-side on the same surface. This finding enriches our understandings of the variant HLBHL growth mechanism and allows one to design various phosphorene architectures by epitaxy for different purposes.

Deposition of P and subsequent STM examinations were carried out in a Unisoku UHV system consisted of a sample preparation chamber and an insert of STM facility operated at 77 K. The base pressure was $1 \times 10^{-10}$ Torr. Crystalline Au(111) was cleaned by $Ar^+$ bombardment (1.0 keV, $5 \times 10^{-6}$ Torr) followed by annealing at ~ 600 °C. Phosphorus vapor was generated from thermal decomposition of InP in a standard Knudsen cell operated at ~ 470 °C. The temperature of the substrate was 210 - 230 °C. After depositing for a preset period, the sample was immediately transferred to the cooling stage of the STM for surface examinations. Electrochemically etched tungsten tips were used, which were conditioned by heating and/or silver-decoration. For all the STM measurement, constant-current mode was used and the tunneling current was 100 pA.

To aid the experiments, atom adsorption geometry and energy of P on Au(111) were investigated using density functional theory (DFT) calculations. The electronic exchange correlation was treated using the Perdew-Burke-Ernzerhof (PBE) form [23] of the generalized gradient approximation (GGA). The projected augmented wave method [24,25] was employed as implemented in the Vienna ab initio simulation package (VASP) [26]. The atom position was optimized by conjugate gradient method until the forces on each atom were smaller than 0.02 eV/Å. The cutoff energy was 500 eV and the vacuum region was set at ~ 12 Å. The Gamma-centered sampling with $6 \times 8 \times 1$ (for $4\sqrt{3} \times 4\sqrt{3}$ supercell), $10 \times 10 \times 1$ (for $3\sqrt{3} \times 3\sqrt{3}$ supercell) or $15 \times 15 \times 1$ (for $3 \times 3$ supercell) $k$ meshes were used. We employed the DFT-D3 [27] method of Grimme to evaluate the van der Waals (vdW) effect in all calculations. Au (111) slab with 5 atomic layers was adopted as the substrate where the bottom layer was fixed to simulate the bulk.

SL blueP has the graphene-like structure but composed of two atomic layers of P

atoms arranged in the buckled honeycomb lattice structure as illustrated in Figure 1a. The unit cell is highlighted by the dashed blue rhombus and the basis vectors $\vec{a}_1$ and $\vec{a}_2$ are along the zigzag edges with the lattice constant $a_1 = a_2 = 0.328$ nm [17]. The buckling of blueP defined as distance between the two P atomic layers is $d \sim 0.13$ nm [28,29]. Figure 1b shows a typical reflected high energy electron diffraction (RHEED) pattern of an epitaxial P layer on Au(111), where the diffraction streaks of Au(111)-1 × 1 and P are clearly resolved as marked by the red and blue arrows, respectively. Fractional streaks (highlighted by the green dashed lines) are also discernable, suggesting a (4 × 4) superstructure of P that is commensurate with the (5 × 5) supercell of Au(111) surface. Therefore, the epitaxial P layer has a lattice constant of ~ 0.35 nm (*i.e.*, $5a_{Au}/4$, where $a_{Au} \sim 0.288$ nm is the lattice constant of Au). This is considerably larger than the theoretical lattice constant of a free-standing blueP (0.328 nm). So, it would be tensile strained by as much as ~ 7% if the epilayer were a blueP SL. Figure 1c shows a STM topographic image of the grown P layer, which reveals regularly arranged dark dots in a triangular network (refer to Supplementary S1 for an image at a larger length scale). The spatial separation between the neighboring dark dots is exactly five times that of Au(111) lattice and thus corresponds well with the RHEED observation. Figure 1d is a close-up image of the same surface, showing a hexapetalous-flower-like structure where each dark core is surrounded by six triangular bright petals. A similar structure has been reported in Ref. [18,20] and was assigned to be an epitaxial blueP SL. We note that such a reconstructed surface of P-on-Au is quite stable and persistent and the line profile measurements (see below) indeed suggest it to be a P double-layer showing an inter-layer distance similar to the buckling height of a blueP SL. Therefore we shall refer to it as blueP or blueP-like layer hereafter. We however would like to remark that the large tensile strain (7%) and the (5 × 5) reconstruction may imply at least a nonideal blueP, which invites further investigations.

The main and most intriguing finding in this work is the unusual kinetic pathway with which the blueP layer is grown. At low coverage, P atoms are seen to aggregate into 1D chains running in three equivalent directions 120° apart as exemplified in Figure 2a. These 1D chains show the bright contrasts with an apparent height of 60 pm when measured at the sample bias of 0.1 V. We remark that this value does not necessarily reflect the geometrical height of P adatoms above the substrate surface due to the factor of density of states (DOS). Nevertheless, it is much shorter than the thickness of a blue or black phosphorene. Moreover, the separation between nearest-neighbor bright spots in the chains is ~ 0.5 nm, corresponding to $\sqrt{3}a_{Au}$, and is inconsistent with that of a phosphorene layer. Besides 1D chains, one also finds one or two small clusters of triangular or truncated-triangular shape that are attached to the end or side of the chain. Many more chains are however seen to simply branch out or intersect with each other, giving rise to the 'Y' shaped structure. Mysteriously, star-shaped intersections are rarely found, reflecting some kind of asymmetry or anisotropy in the chain formation kinetics. As the dosage of P increases, more and enlarged compact clusters/islands (similar to Figure 2d) are found and in rare cases large areas of the surface become covered by compactly arranged spots with $\sqrt{3}a_{Au}$ lateral separations, resulting in the $(\sqrt{3} \times \sqrt{3})R30°$ reconstructed surface (Figure 2b).

The 1D chain structure of P on Au(111) has not been reported before. We note however a similar chain structure being reported for sulfur (S) adsorption on Au (111), where a repulsive pair interaction but attractive trio-interaction of S adatoms were found responsible for the chain formation [30]. For P adsorption on GaN(0001) surface, a repulsive adatom interaction was also suggested [19]. Our own DFT calculations of P adsorbed on Au(111) show a more complicated behavior. At low coverage (≤ 1/9 monolayer (ML), where 1 ML is defined as one P adatom per Au(111) surface site), isolated P monomers adsorbed at the fcc sites of Au(111) is more stable than the chemically bonded P-trimers. On the other hand, P monomers tend to be closely spaced by occupying the second nearest-neighbor fcc sites with inter-atom distances $\sqrt{3}a_{Au}$ apart (refer to Supplementary S2). The energy gain for the latter amounts to ~ 60 meV/adatom. Interestingly, it is also found that at even lower coverage (≤ 1/12 ML), the chain configuration is more favorable than the close-packed one by 18 meV/adatom (see Supplementary S2), which explains the 1D-chains seen for the low-coverage surface of Figure 2a. Indeed, the average chain-chain separation in Fig. 2a is about 2 nm or $7a_{Au}$, which translates into a coverage of $1/(7 \times \sqrt{3}) \approx 1/12$ ML. Such a configuration change may reflect an effect of P-adatom, chain and substrate interaction via structural relaxation or charge transfer effect, which could be appreciable in the P-on-Au system. With increasing P coverage (≥ 1/9 ML), the energy difference between the chain and compact configurations becomes minimal and so one expects the chain structure to gradually evolve into more compact $R30°$ patches and finally to one of the whole surface being covered by the P-adatom induced $(\sqrt{3} \times \sqrt{3})R30°$ at the coverage of 1/3 ML and the P-trimers at 1 ML before blueP islands nucleate and grow at coverages above 1 ML.

It is however surprising that we seldom capture surfaces showing compact $(\sqrt{3} \times \sqrt{3})R30°$ or P-trimers over large areas in experiment. Instead, we often observe blueP islands nucleated in the background of loose $(\sqrt{3} \times \sqrt{3})R30°$ patches or 1D chains. An example is shown in Figure 2c and an enlarged image of the surrounding area is given in Figure 2d. The hexagonal island shows the hexapetalous-flower-like structure on top, characterizing the blueP-like layer as mentioned earlier. The inset shows a line profile across the island step, revealing a height of ~ 0.15 nm, which is very close to the buckling height of a blue phosphorene

SL (i.e., 0.13 nm when measured from the top to bottom layer P atoms, refer to Fig. 1a). So such an island appears to be composed of a P double-layer with the interlayer distance consistent with blueP. However, as is clear from Figs. 2c and 2d, such blueP islands appear to have nucleated out of the 1D P-chains or loose $(\sqrt{3} \times \sqrt{3})R30°$ background rather than compact $(\sqrt{3} \times \sqrt{3})R30°$ or P-trimer surfaces.

Figures 3a presents another STM image of a sample at a slightly higher coverage and Figs. 3b and 3c show line profile measurements along respectively the horizontal white and vertical blue dashed lines drawn in Fig. 3a. Again, they reveal the apparent height of ~ 0.15 nm for the blueP islands (patch) as well as the step height of Au(111) surface (~ 0.24 nm). In Figure 3a, the large blueP patch has some unfilled holes in the middle, which also expose the loose $(\sqrt{3} \times \sqrt{3})R30°$ structure similar to that seen in the top region of the same image. The line profile of Fig. 3c taken across one such unfilled holes and extended to the top region reveals that they are indeed of the same height and is 0.15 nm below the blueP surface. Fig. 3b taken along a line crossing a step of the substrate and with the same loose $(\sqrt{3} \times \sqrt{3})R30°$ structure at the far left and far right regions but a blueP patch in the middle shows two steps of heights 0.09 nm and 0.15 nm, which add up to give rise to the total step height of 0.24 nm, matching exactly that of a monatomic step on the Au (111) surface.

Returning to the question why compact $(\sqrt{3} \times \sqrt{3})R30°$ and P-trimer surfaces are rarely observed by experiments, we suggest it is related to the reduction of system energy by balancing different configurations under a fixed total coverage. For example, the surface of compact $(\sqrt{3} \times \sqrt{3})R30°$ or P-trimer at coverage of 1/3 or 1 ML may be broken into a composite surface consisted of 1D-chains with low local coverage and blueP islands with high local P coverage, yet the overall coverage is maintained. Our total energy calculations show that the compact $(\sqrt{3} \times \sqrt{3})R30°$ and P-trimer surfaces are indeed less favorable than the 1D-chain or blueP configurations. Figure 4 presents the calculated formation energies for the 1D-chain, compact $(\sqrt{3} \times \sqrt{3})R30°$, P-trimer and blueP configurations, which are stable configurations at the respective coverages (refer to Supplementary S2). As seen, the $(\sqrt{3} \times \sqrt{3})R30°$ and P-trimer surfaces both have higher formation energies than the 1D-chain and blueP structures. Therefore, as P adatom coverage builds up, instead of forming compact $(\sqrt{3} \times \sqrt{3})R30°$ or P-trimer structures over large areas, local fluctuation may cause blueP nucleation in certain places, consuming some P adatoms from the already formed but small $(\sqrt{3} \times \sqrt{3})R30°$ patches and in return converting the latter into the 1D-chains or the loose $(\sqrt{3} \times \sqrt{3})R30°$ configuration. This may be viewed as a phase separation of an otherwise homogeneous $(\sqrt{3} \times \sqrt{3})R30°$ or P-trimer structured surface and the growth process thus no longer follows the sequential chain, $(\sqrt{3} \times \sqrt{3})R30°$, and P-timer formation before nucleation of blueP.

The above finding firstly lends an experimental support to the recent theoretical proposal of the HLBHL growth mechanism [19]. More importantly, it offers kinetic and structural details. The $(\sqrt{3} \times \sqrt{3})R30°$ and P-trimer surfaces would represent the first half-layer but before they practically realize over large areas, local fluctuations lead to nucleation of blueP in some places by consuming P adatoms in the nearby regions, giving rise to a composite surface of locally lower-coverage 1D-chains surrounding the high-coverage blueP islands. This is exemplified better in Fig. 4b.

To conclude, we have demonstrated P growth on Au(111) by MBE and discovered a new 1D-chain structure of P adatoms at low coverage. The 1D chains evolve into $(\sqrt{3} \times \sqrt{3})R30°$ patches before blueP-like islands nucleate. However, instead of the homogeneous $(\sqrt{3} \times \sqrt{3})R30°$ or P-trimer surface being fully grown, our experiments and DFT calculations suggest a 'phase-separated' configuration where the surface contains the coexisting 1D-chains and blueP islands having local low and high P coverages, respectively. Yet the overall coverage is maintained. This finding thus supplements an early theoretical proposal of the novel HLBHL growth mechanism of blue phosphorene with kinetic and structural details.

**Acknowledgements**

This work was financially supported by a Collaborative Research Fund (HKU9/CRF/13G) sponsored by the Research Grants Council (RGC), Hong Kong Special Administrative Region.

Jin-Peng Xu, Jun-Qiu Zhang and Hao Tian contributed equally to this Letter.

**Figures and Captions**

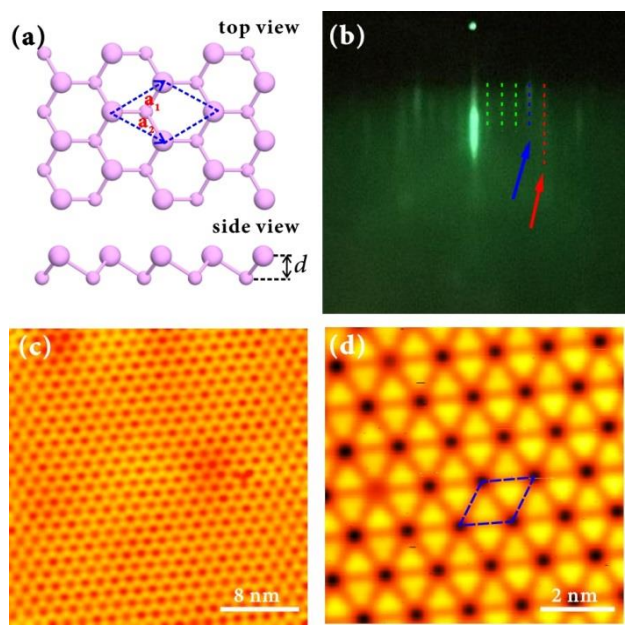

**Figure 1**. Blue phosphorene on Au(111). (a) Top-view (upper) and side-view (lower) of a model of blueP layer. The blue rhombus drawn in the upper panel represents the unit cell. (b) RHEED pattern of epitaxial phosphorene on Au(111). The red and blue arrows point the 1 × 1 diffraction streaks of Au substrate and epitaxial P, respectively, and the green dashed lines mark the fraction streaks due to the (5 × 5) reconstruction of epitaxial P layer. (c) STM topographic image of blueP-like surface (size: 30 × 30 nm$^2$, sample bias $V_{bias}$ = 1 V). (d) A close up image of the same surface (Size: 8 × 8 nm$^2$, $V_{bias}$ = 0.1 V).

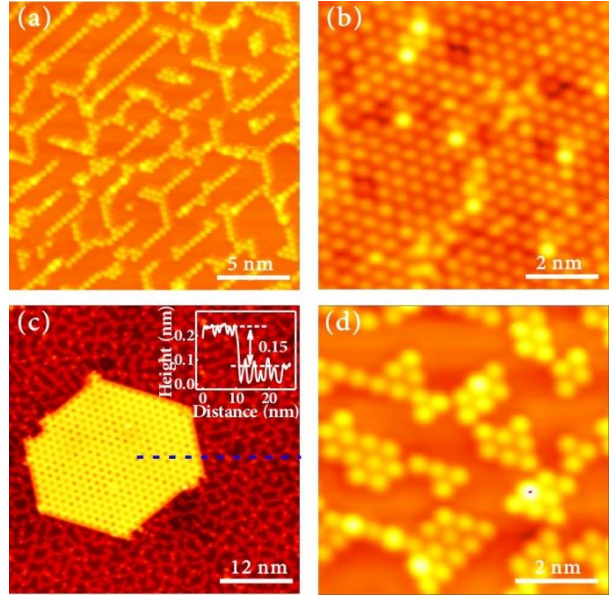

**Figure 2**. Growth sequence of epitaxial blueP. (a) STM image (Size: 20 × 20 nm², $V_{bias}$ = 0.1 V) of a surface after exposing Au(111) surface by a low dosage of P, revealing the 1D chain structure with spot separation of $\sqrt{3}a_{Au}$. (b) STM image (Size: 8 × 8 nm², $V_{bias}$ = 0.1 V) of the surface at higher P dosage, showing the compact close-packed $(\sqrt{3} \times \sqrt{3})R30°$ structure. (c) STM image (Size: 50 × 50 nm², $V_{bias}$ = 0.1 V) of an isolated blueP-like island surrounded by the the 1D chains and loose $(\sqrt{3} \times \sqrt{3})R30°$ patches in the background. The inset shows a line profile along the dashed horizontal line. (d) A zoom-in STM image (Size: 7 × 7 nm², $V_{bias}$ = 0.1 V) of the surrounding region of the blueP island in (c).

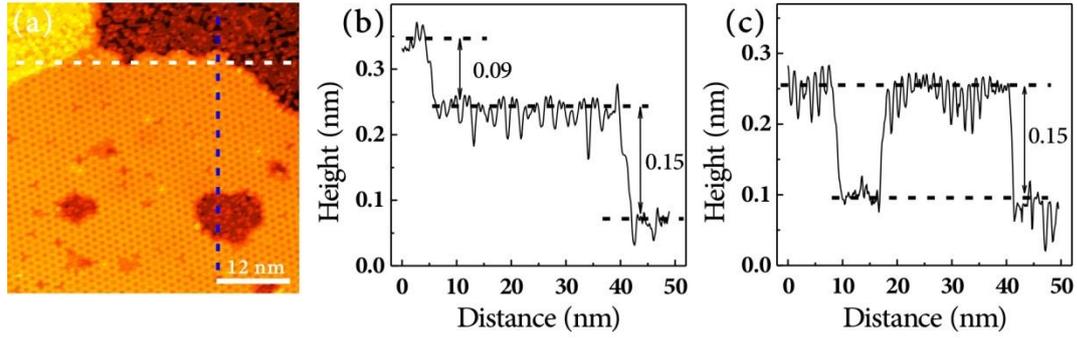

**Figure 3**. Buckling height of blueP. (a) STM image (Size: 50 × 50 nm$^2$, $V_{\text{bias}}$ = 0.1 V) of a surface containing mixed regions of $(\sqrt{3} \times \sqrt{3})R30°$ and a blueP terrace. Note also the $(\sqrt{3} \times \sqrt{3})R30°$-structured hole within the blueP terrace. (b, c) Line profiles taken along the horizontal white (b) and vertical blue (c) dashed lines in (a), revealing the step heights of various kinds.

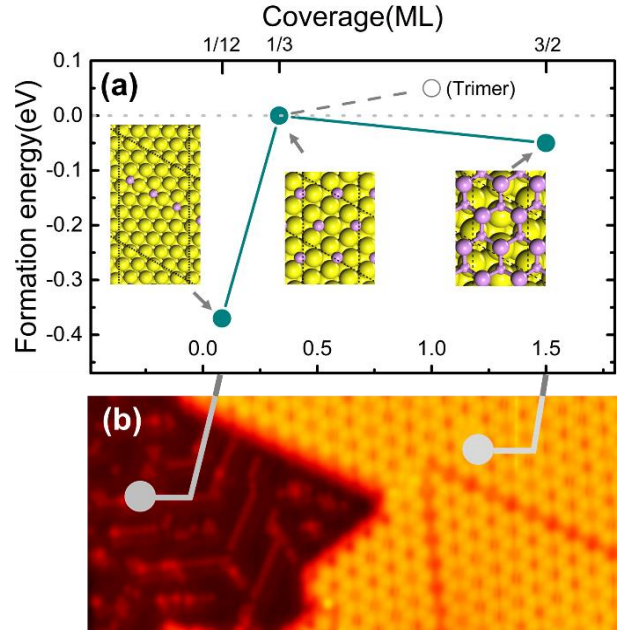

**Figure 4**. Phase-separated composite surface. (a) Relative formation energies per P adatom of some stable configurations (insets) of P-on-Au(111), i.e., 1D chain, compact $(\sqrt{3} \times \sqrt{3})R30°$, P-trimer and blueP, at different coverages. The formation energy $E_f$ is defined as $E_f = (E_t - E_{sub})/N$, where $N$ is the number of P adatoms, $E_t$ is the total energy and $E_{sub}$ is that of the substrate. The $E_f$ of the $(\sqrt{3} \times \sqrt{3})R30°$ structure is set at zero. The yellow and purple balls in the inset represent Au and P atoms, respectively. For P atoms in blueP, the ball size marks the vertical position where the larger balls represent higher positions. (b) STM image (Size: 31 × 14 nm$^2$, $V_{bias}$ = 0.1 V) showing locally low-coverage 1D-chains (left) surrounding the high-coverage blueP island/terrace (right).

# Supplementary Information

# One-dimensional phosphorus chain and two-dimensional blue phosphorene grown on Au(111) by molecular-beam epitaxy

Jin-Peng Xu[1], Jun-Qiu Zhang [1], Hao Tian[1,2], Hu Xu[2], Wingkin Ho[1] and Maohai Xie[1*]

[1]*Department of Physics, Hong Kong University, Hong Kong, China*
[2]*Department of Physics, Southern University of Science and Technology, Shenzhen, 518055, China*

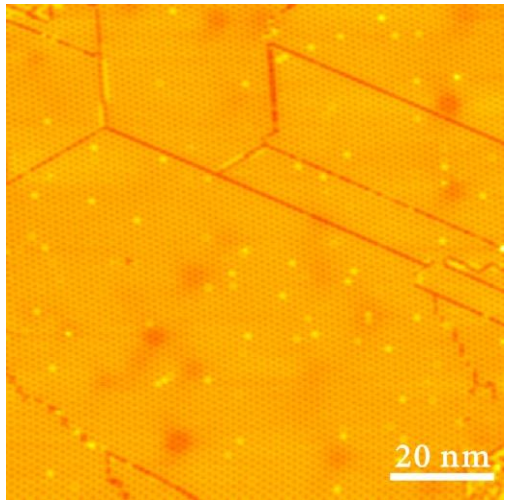

Figure S1. A large scale STM image of buleP on Au(111). Size: 100 × 100 nm$^2$, $V_{bias}$ = 1 V, I = 100 pA.

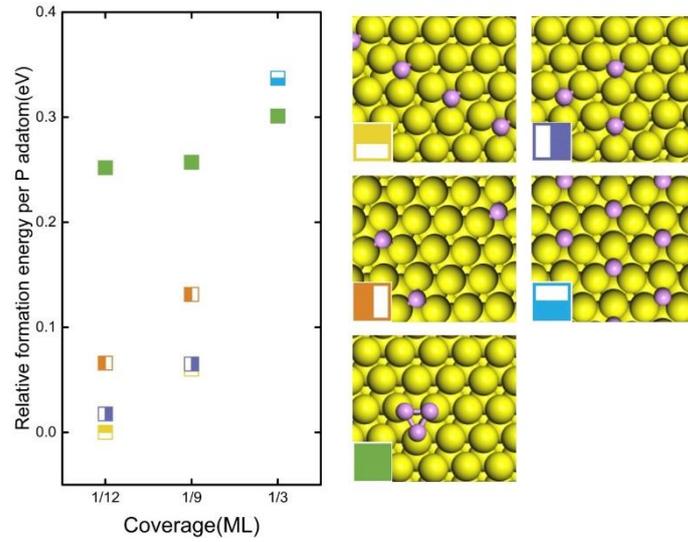

Figure S2. Relative formation energy per P adatom of various configurations (right panels) at the nominal coverage of 1/12, 1/9 and 1/3 MLs. Here, the formation energy is defined as $E_f = (E_t - E_{sub})/N$, where $N$ is the number of P adatoms, $E_t$ is the total energy of each configuration, and $E_{sub}$ is the energy of the substrate. The $E_f$ of the chain structure at 1/12 ML is set as zero, while the energy difference between the various configurations with the chain structure are calculated and plotted. The solid square denotes trimmer configuration and the half-solid symbols denote single adatoms in different arrangements.